\documentclass{kapproc} 

\usepackage{t1enc}

\usepackage{procps} 

\usepackage[dvips]{graphicx}

\upperandlowercase

\kluwerbib 

\begin{document}

\articletitle[]
{Optical and near-infrared properties of submillimetre galaxies in the GOODS-NOrth field
}
\author{Alexandra Pope,\altaffilmark{1} Douglas Scott,\altaffilmark{1} Colin Borys,\altaffilmark{2} Christopher Conselice,\altaffilmark{2} Mark Dickinson,\altaffilmark{3,4} Bahram Mobasher\altaffilmark{4}}

\affil{\altaffilmark{1} University of British Columbia, Vancouver, Canada, \ 
\altaffilmark{2} California Institute of Technology, Pasadena, USA, \ \altaffilmark{3} National Optical Astronomy Observatory, Tucson, USA, \ \altaffilmark{4}Space Telescope Science Institute, Baltimore, USA
}

\begin{abstract}
We investigate the photometric redshift distribution, colours and morphologies of a large sample of submillimetre sources in the GOODS-North region. With the depth achieved by the deep ACS images, optical counterparts have been found for all the radio-detected sub-mm sources, and we have used several techniques to identify counterparts to the radio-undetected sources. The sub-mm counterparts have a median photometric redshift, from optical and near-IR photometry, of 1.9. Our sample contains a much higher fraction of EROs than other sub-mm surveys, and larger angular sizes than the average galaxy at all redshifts. The sub-mm galaxy counterparts are consistent with massive star-forming galaxies at high redshift. 
\end{abstract}


\section{Introduction}
Extragalactic submillimetre (sub-mm) surveys have revealed a population of high redshift galaxies significant for early galaxy formation and evolution and which are thought to be the progenitors of present-day massive elliptical galaxies (Lilly et al.~1999, Blain et al.~2002, Scott et al.~2002). Progress in understanding this population is made by studying the characteristics of individual sources at other wavelengths. This is challenging due to the large JCMT beam size and faint nature of these sources in optical images. Coincidence with a radio source is a successful way to identify the counterpart to the sub-mm emission (Barger, Cowie, \& Richards 2000; Ivison et al.~2002; Smail et al.~2002; Borys et al.~2004; Clements et al.~2004), and then attempt to measure spectroscopic redshifts (Chapman et al.~2003). However, SCUBA sources detected at $1.4\,$GHz represent only about half of the total number of sources found in all extragalactic SCUBA surveys, and therefore there is a substantial fraction of SCUBA sources about which we currently know very little. 

The Great Observatories Origins Deep Survey (GOODS,Giavalisco et al.~2004) provides an ideal data--set for studying sub-mm sources as it combines the deepest observations across all wavelengths. We present a study of a large sample of SCUBA galaxies in the GOODS-North region using the ACS HST optical images and ground-based near-IR observations. Our aim is to understand the whole sub-mm population, including both the radio-detected and radio-undetected sub-samples.

\section{Submillimetre sample}
The largest amount of blank-field SCUBA data in a single field is found in the HDF-N region. Data from about 70 shifts of telescope time have been carefully combined into a `super-map' of the region (see Borys et al.~2003 and references therein). In addition to the data presented in Borys et al.~2003, we have obtained new data and our updated source list contains 22 objects at $>4\sigma$ with an additional 18 at  $3.5$--$4.0\sigma$, all but one of which overlap with the deep $HST$ observations from GOODS. 

We have identified unique optical counterparts for 15 radio-detected and 13 radio-undetected sub-mm sources (Pope et al.~2005). The deep ACS images have revealed many new optical counterparts, not present in shallower surveys. Half of our optical counterparts would not have been detected in a typical SCUBA follow-up image with a $i_{\rm{775}}$ magnitude limit of $\sim25$.

\section{Redshift distribution}

\begin{figure}[ht]
\includegraphics[width=4.7in,angle=0]{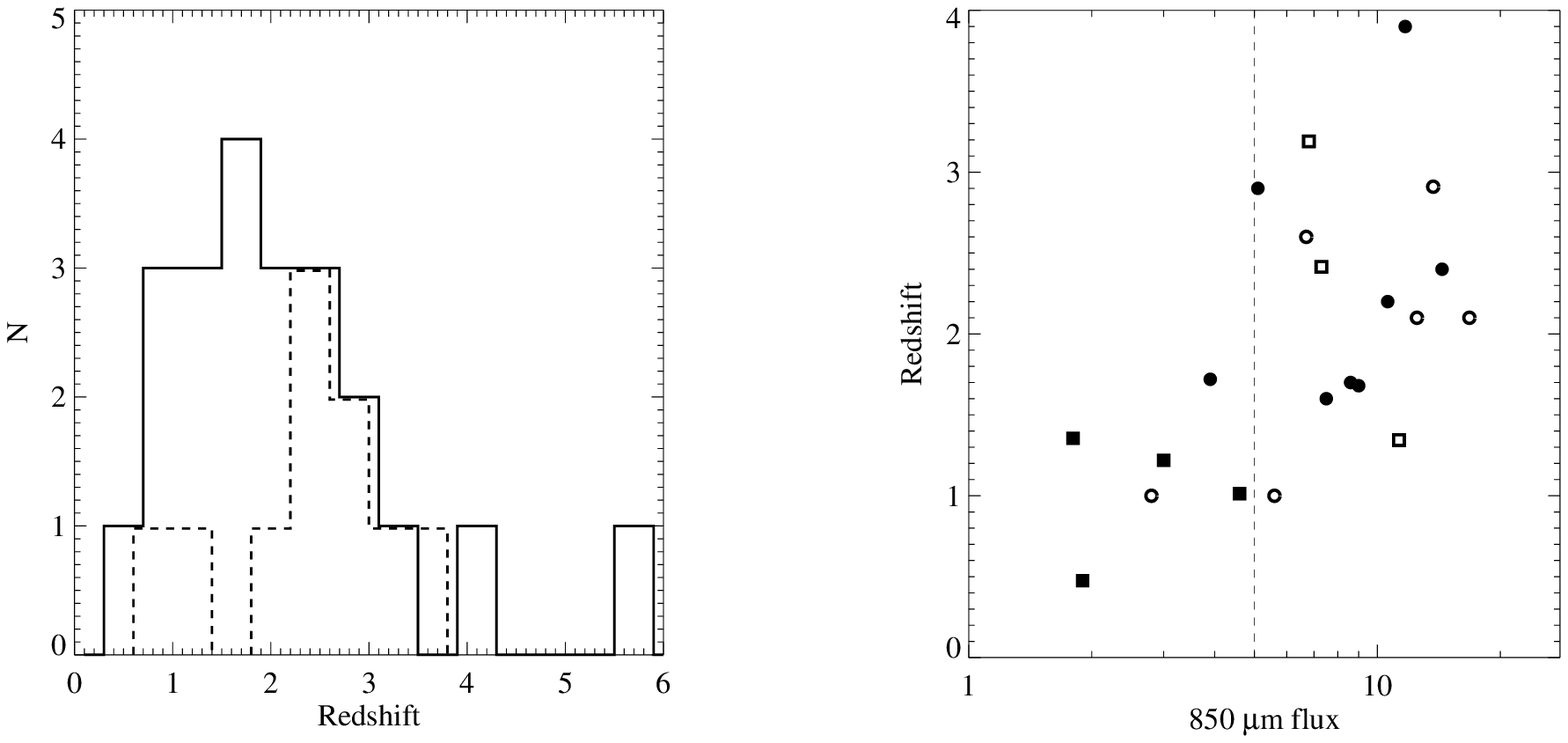}
\sidebyside
{\caption{Redshift distribution for sub-mm sources. The photometric redshifts for our full sample is plotted as the solid line. The dashed line is the distribution from Chapman et al. (2003) of radio-selected sub-mm-bright sources. 
}}
{\caption{Redshift as a function of 850$\,\mu$m flux density on a logarithmic scale. The solid symbols are the radio-detected sub-mm sources while the open symbols are the radio-undetected sub-mm sources. The squares denote sources which have spectroscopic redshifts, while the circles have photometric redshift estimates. 
}}
\end{figure}
\noindent

Photometric redshifts have been estimated for a large sample of sources in GOODS-North using both a $\chi^{2}$ minimization technique (Puschell, Owen, \& Laing 1982) and a Bayesian method (Ben{\'{\i}}tez 2000).

In Fig. 1, we plot the distribution of redshifts for all our sub-mm sources, which has a median redshift of 1.9 and a quartile range of 1.3--2.6. For comparison, we have also plotted the distribution from spectroscopic studies. Chapman et al.~(2003) have targetted a sample of sub-mm bright radio-detected sources and find a median redshift of 2.4. Our median value is slightly lower, which may be because photometric redshift estimates are not affected by the `redshift-desert' at $z\simeq1.5$ apparent in the Chapman et al.~2003 distribution. Since we are not constrained by the radio observations, we might expect our full distribution to go out to higher redshifts, but we find no evidence for such a high redshift tail. We also find that the radio-undetected sources have an only marginally higher redshift distribution (Pope et al.~2005). This suggests that the difference is dominated by scatter about the radio threshold, and that almost all SCUBA galaxies would be detected in only moderately deeper radio maps.

Another intriguing result shown in Fig.~2 is that there appears to be a lack of faint ($S_{850}\sim3\,$mJy) SCUBA galaxies at high redshift. Recall that because of the negative K-correction at 850$\,\mu$m, the observed flux density for a galaxy with a specific luminosity is essentially constant past $z\sim1$, meaning that fainter SCUBA sources are typically intrinsically less luminous (Blain et al.~2002). Although there may be several other explanations, the obvious implication is that at high redshift the SCUBA galaxies form a distinct population, i.e. major mergers as opposed to more quiescent star-forming galaxies.

\section{Colours}

\begin{figure}[ht]
\begin{center}
\includegraphics[width=4.7in,angle=0]{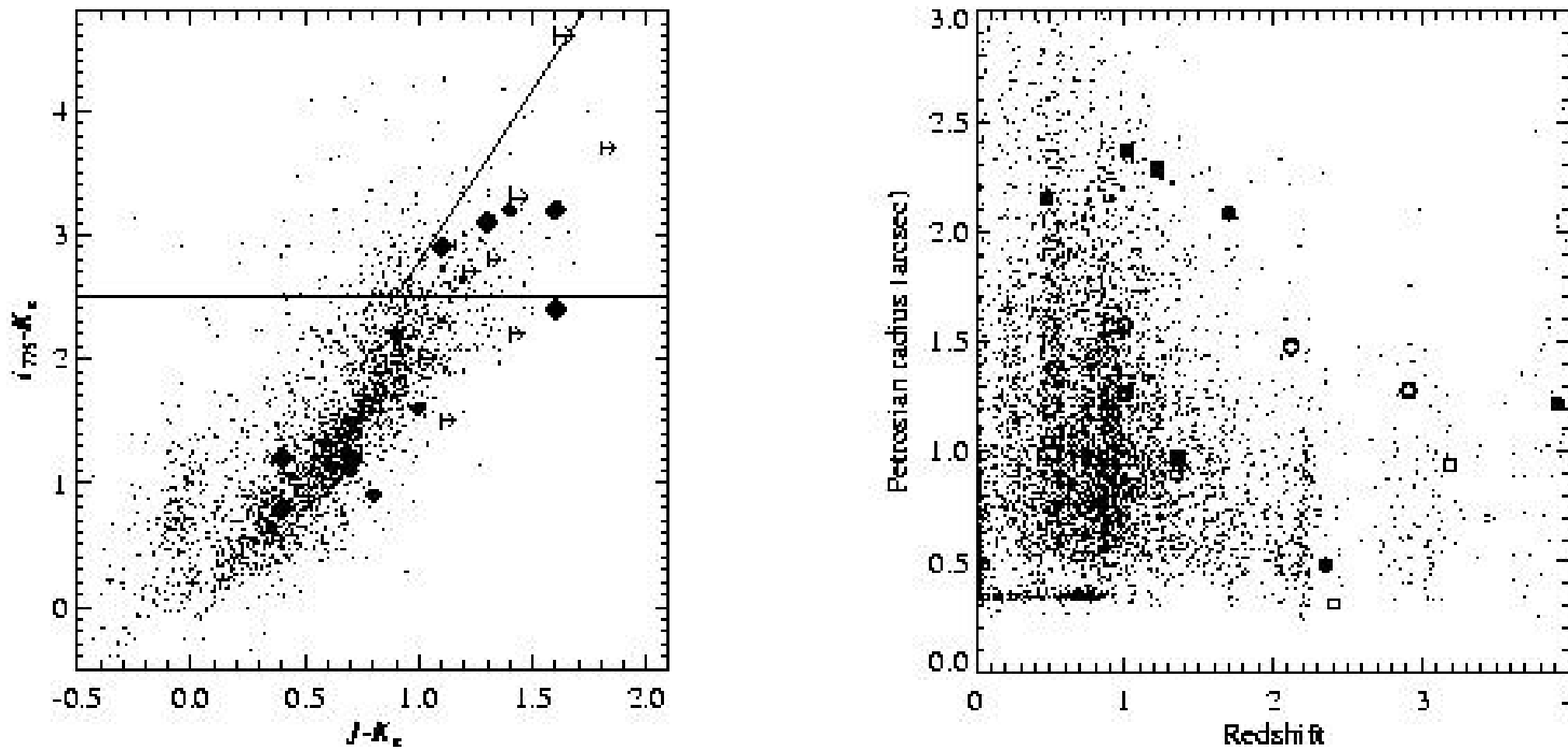}
\sidebyside
{\caption{Near-IR colour-colour plot. Dots are field galaxies from the GOODS catalogue. The large symbols are the sub-mm counterparts with $1<z<2$. Pozzetti \& Mannucci (2000) state that, within this redshift range, sources which are starburst EROs, as opposed to ellipticals at $z\simeq1$, are expected to fall to the right of the diagonal line in this colour-colour plot. The horizontal line is the the traditional ERO cut-off through these filters.}
}
{\caption{Sizes of SCUBA galaxies. The Petrosian radius is plotted as a function of redshift. Solid symbols are the radio-detected sub-mm sources and the open symbols are the radio-undetected sub-mm sources. Sources with spectroscopic redshifts are plotted as squares and those with photometric redshifts are denoted by circles. }
}
\end{center}
\end{figure}
\noindent

An ERO as defined through our filters corresponds to $(i_{\rm{775}}-K_{\rm{s}})_{\rm{AB}}>2.5$. With this criterion, 65 per cent of the radio-detected sub-mm sources with a near-IR counterpart are EROs. This is a much higher ERO fraction for the radio-detected counterparts than previous SCUBA surveys, which is due to the depth of the ACS data. Pozzetti \& Mannucci (2000) discuss the two main types of EROs, namely elliptical galaxies at $z\sim1$ and dusty starburst galaxies. When plotted in a colour-colour diagram, the EROs with $1<z<2$ are clearly a separate population. In Fig.~3, we show the colour-colour plot for the ACS field galaxies and the sub-mm counterparts. We find that all of our sub-mm sources which are EROs at $1<z<2$ are consistent with being starburst galaxies.

\section{Morphologies}

To investigate if sub-mm galaxies are indeed evolving into massive ellipticals, we can study their morphologies. One of the main advantages of the deep ACS images is the high resolution, which allows us to study the structure of the sub-mm counterparts. Fig.~4 shows how the sizes of sub-mm counterparts compare to other high-redshift galaxies. We see that the Petrosian radius of the sub-mm galaxies is larger than for ACS field galaxies at essentially all redshifts. 

\section{Summary}

Using the optical and near-IR images from GOODS, we have identified and characterized a large fraction of sub-mm sources. With the deep optical and near-IR data, we estimate photometric redshifts for our sub-mm sample and find a median redshift of 1.9. The sub-mm galaxies are red both in $i_{\rm{775}}-K_{\rm{s}}$ and $J-K_{\rm{s}}$. These colours together are most useful for selecting and describing the counterparts. The sub-mm counterparts are large galaxies given their redshift. Our current work suggests that optical imaging to $i_{\rm{775}}\simeq28$, coupled with deep radio data, should be sufficient to identify the bulk of the sub-mm galaxies in the future sub-mm surveys.

As part of GOODS, observations with Spitzer of GOODS-North are ongoing. We will use these deep infrared observations to verify counterparts and the additional photometry points will be invaluable in understanding the optical/IR SEDs of SCUBA galaxies.

\end{document}